\documentclass[aps,prd,titlepage,groupedaddress,longbibliography,nofootinbib,preprint,12pt]{revtex4-2}
\usepackage[utf8]{inputenc} 
\usepackage[T1]{fontenc} 
\usepackage{lmodern} 
\usepackage{microtype} 
\usepackage[english]{babel}
\usepackage{color}
\usepackage{graphicx}
\usepackage{amsmath}
\usepackage{amssymb}
\usepackage{mathrsfs}
\usepackage{slashed}
\usepackage{dsfont}
\usepackage[usenames,dvipsnames]{xcolor}
\usepackage[colorlinks=true, linkcolor=black, citecolor=Mahogany, urlcolor=Fuchsia]{hyperref}
\usepackage{changes}

\usepackage{setspace}
\usepackage{etoolbox}
\makeatletter
\patchcmd{\@footnotetext}
  {\setspace@singlespace}{1}
  {}{}
\makeatother

\usepackage{fancyhdr}
\def\be{\begin{equation}}
	\def\ee{\end{equation}}
\def\bes{\begin{equation*}}
	\def\ees{\end{equation*}}
\def\bea{\begin{eqnarray}}
	\def\eea{\end{eqnarray}}
\def\bal{\begin{align}}
	\def\eal{\end{align}}

\def\f{\frac}
\def\mc{\mathcal}

\def\w[#1]{\widehat{#1}}
\def\vs[#1,#2]{\boldsymbol{{#1}_{#2}}}
\def\mes[#1]{d^{3}{#1}}
\def\del{\partial}
\def\<{\langle}
\def\>{\rangle}
\def\vecs[#1,#2]{\boldsymbol{{#1}_{#2}}}

\def\a{\alpha}
\def\b{\beta}
\def\d{\delta}
\def\D{\Delta}
\def\e{\epsilon}

\def\g{\gamma}
\def\G{\Gamma}
\def\k{\kappa}
\def\l{\lambda}
\def\L{\Lambda}
\def\m{\mu}
\def\n{\nu}
\def\N{\nabla}
\def\o{\omega}
\def\O{\Omega}

\def\P{\Phi}

\def\s{\sigma}

\def\z{\zeta}

\begin{document}

\mbox{}\vspace{10mm}

\title{Beyond Drude transport in hydrodynamic metals}

\author{Blaise Gout\'eraux}
\email{blaise.gouteraux@polytechnique.edu}
\affiliation{CPHT, CNRS, \'Ecole polytechnique, Institut Polytechnique de Paris, 91120 Palaiseau, France}
\author{Ashish Shukla}
\email{ashish.shukla@polytechnique.edu}
\affiliation{CPHT, CNRS, \'Ecole polytechnique, Institut Polytechnique de Paris, 91120 Palaiseau, France}

\preprint{CPHT-RR052.072023}

\setstretch{1.26}

\begin{abstract}
In interacting theories, hydrodynamics describes the universal behavior of states close to local thermal equilibrium at late times and long distances in a gradient expansion. In the hydrodynamic regime of metals, momentum relaxes slowly with a rate $\Gamma$, which formally appears on the right-hand side of the momentum dynamical equation and causes a Drude-like peak in the frequency dependence of the thermoelectric conductivities. Here we study the structure and determine the physical implications of momentum-relaxing gradient corrections beyond Drude, \emph{i.e.} arising at subleading order in the gradient expansion. We find that they effectively renormalize the weight of the Drude pole in the thermoelectric conductivities, and contribute to the dc conductivities at the same order as previously-known gradient corrections of translation-invariant hydrodynamics. Turning on a magnetic field, extra derivative corrections appear and renormalize the cyclotron frequency and the Hall conductivity. This relaxed hydrodynamics provides a field-theoretic explanation for previous results obtained using gauge-gravity duality. In strongly-coupled metals where quasiparticles are short-lived and which may be close to a hydrodynamic regime, the extra contributions we discuss are essential to interpret experimentally measured magneto-thermoelectric conductivities. Specializing to metals close to a Fermi liquid phase, the effective mass measured either through the specific heat or the spectral weight of the Drude-like peak are found to differ, as was indeed reported in overdoped cuprate superconductors. More generally, we expect such terms to be present in any hydrodynamic theory with approximate symmetries, which arise in many physical systems.
\end{abstract}

\maketitle

\setstretch{1.2}

\tableofcontents

\section{Introduction and Summary of results}
In interacting systems, hydrodynamics provides a universal framework capturing the effective dynamics of states slowly varying about local thermal equilibrium, \cite{landaubook,forster,chaikinlubensky1995,Kovtun:2012rj}. Besides classical liquids, a non-trivial example of hydrodynamic flow is found in the quark-gluon plasma produced in heavy ion collisions. On the other end of the energy spectrum, its applicability to metallic phases in condensed matter systems was traditionally limited by the fact that in a typical metal, inelastic collisions of electrons with lattice phonons or impurities happen on scales comparable to elastic electron-electron collisions. The latter tend to restore local equilibrium, while the former degrade momentum at a fast rate, invalidating the assumptions underlying hydrodynamics. In recent years, thanks to ultra pure samples becoming available, Graphene has now become an important platform for the investigation of hydrodynamic effects in metals, see \cite{Lucas:2017idv,fritz2023hydrodynamic} for reviews. However, hydrodynamics in the sense of an effective theory of states with conserved energy, momentum and number density does not apply stricto sensu, as inelastic scattering due to disorder or umklapp relaxes momentum. Yet, if such inelastic scattering is weak enough, momentum relaxes slowly and can be treated as an approximate symmetry.  

To leading order, one simply adds a relaxation rate $\Gamma$ to the momentum (non-)conservation equation \eqref{MomRelaxEq}, see e.g.~\cite{Hartnoll:2007ih}. This captures the dominant effects of weakly-broken spatial translations, namely the appearance of a purely imaginary mode with a parametrically small gap $\omega\simeq-i\Gamma$ in the spectrum of collective excitations, and the broadening of the momentum-conserving delta function in thermoelectric conductivities into a sharp Drude-like peak. To with, the electric conductivity $\sigma(\omega)$ receives a hydrodynamic (`non-Ohmic') contribution due to the dynamics of (approximately-)conserved momentum: 
\begin{equation}
\sigma(\omega)=\frac{K_D}{(-i\omega)}+\dots\quad\underset{\Gamma\neq0}{\longrightarrow}\quad \frac{K_D}{\Gamma-i\omega}+\dots
\end{equation}
featuring the Drude-like pole mentioned above with spectral weight $K_D$. The dots stand for analytic contributions, see \eqref{acsigma} below for a more precise expression.

The momentum relaxation rate $\Gamma$ is a dissipative parameter, as can also be seen by checking that it causes entropy production. Using the memory matrix approach \cite{forster,Hartnoll:2016apf}, it can be computed in terms of the two-point function of the operator $O$ breaking translations (which need not be a scalar operator), with typically $\Gamma\sim (\partial\Phi)^2$ where $\Phi$ is the coupling to $O$. In the spirit of effective field theories, the scaling $\Gamma\sim (\partial\Phi)^2\sim \partial_t\sim\partial_i\sim\ell^2$ allows to probe the dynamics of slowly-relaxing momentum, \cite{Davison:2015bea,Blake:2015epa,Blake:2015hxa}, where we have introduced the scaling parameter $\ell$ for convenience. Importantly, $\ell^2\ll \Lambda_{UV}$, the cut-off scale of the effective theory, which is typically set by temperature, $\Lambda_{UV}\sim T$. In this regime, momentum relaxes on distances large compared to the local equilibration length (eg the mean free path in kinetic theory). {When hydrodynamics is derived from electronic kinetic theory, then $\Lambda_{UV}$ is controlled by $1/\tau_{ee}$, the typical (elastic) electron-electron relaxation rate. On the other hand, $\ell$ is controlled by the electron-phonon $\ell_{e-ph}$ or electron-impurity $\ell_{imp}$ scattering lengths. Thus our theory will be able to describe the crossover between the hydrodynamic regime, on scales $\ell^2\lesssim\omega,k\ll \Lambda_{UV}$, and the diffusive, `ohmic' regime, $\omega,k\lesssim\ell^2\ll \Lambda_{UV}$. It does not capture the crossover to the ballistic regime, where the size of the system is small compared to the electron-electron mean free path and scattering of electrons occurs primarily at the boundary. }

Hydrodynamics allows to treat dissipative effects systematically in a gradient expansion around local thermal equilibrium \cite{Kovtun:2012rj}, leading to e.g.~viscous effects or thermal diffusion in fluids. These effects are parameterized by gradient corrections to the ideal constitutive relations of spatial fluxes, which in turn contribute to second order in gradients to the equations of motion -- thus, at subleading order compared to the leading term causing momentum to relax. This means that a consistent treatment of gradient corrections in momentum-relaxed hydrodynamics must necessarily include subleading translation-breaking gradient corrections as well, or in other words, take into account the next order in perturbation theory around the translation invariant state. 

This is what we set out to do in this work. For concreteness, we focus on breaking spatial translations through a set of spatially-dependent scalar deformations, $O_I$, $I=1\ldots d$, where $d$ is the number of spatial dimensions. As we further elaborate {around \eqref{MomRel} below}, a special choice is when they correspond to the spatially-varying part of the densities of conserved operators, when translations are broken e.g.~by an inhomogeneous chemical potential, \cite{Andreev:2010,Lucas:2015lna,Li:2020,Chagnet:2023xsl}. We will adopt a more general perspective and simply evaluate the expectation values of the $O_I$ in the thermal ensemble order by order in the derivatives of temperature $T$, chemical potential $\mu$, fluid velocity $v^i$ and the scalar couplings $\Phi_I$. Compared to ordinary hydrodynamics, an extra set of gradient corrections appears at the same order as other, more familiar gradient corrections such as the shear viscosity, \cite{Bhattacharyya:2008ji,Ashok:2013jda,Blake:2015epa,Blake:2015hxa,Armas:2021vku,Armas:2022vpf}.

For illustrative purposes and with a widely used gauge-gravity model for a hydrodynamic metal in mind, \cite{Donos:2013eha,Andrade:2013gsa,Donos:2014uba,Gouteraux:2014hca,Davison:2014lua,Davison:2015bea}, we then study linear perturbations around an equilibrium state where $\Phi_I=\bar\Phi_I\equiv\ell\delta_{Ii}x^i$, which preserves isotropy. {$\ell$ is thus the scale which characterizes the strength of momentum relaxation. As we shall see in the next section, $\Gamma\sim\mathcal O(\ell^2)$, so that momentum relaxes on timescales of $\mathcal O(\ell^{-2})$.} We leave to future work the treatment of more realistic inhomogeneous equilibrium states, \cite{Andreev:2010,Lucas:2015lna,Lucas:2017vlc,Li:2020,Chagnet:2023xsl}. In the hydrodynamic regime, they give rise to an effective momentum relaxation rate controlled by first-order gradient corrections and the impurity scattering the typical inhomogeneity length scale, $\Gamma\sim\ell^2\eta$. 

The linearized momentum relaxation equation then features subleading $\mathcal O(\ell^4)$ momentum-relaxing terms $\lambda_{\rho,s,v}$:
\begin{equation}
\label{MomRelaxEq}
\left(1+\f{\ell^2 \l_v}{\chi_{\pi\pi}}\right)\left(\partial_t +\Gamma\right)\pi^i+\partial_j\tau^{ji}=-\ell^2\lambda_s \partial^i T-\ell^2\lambda_\rho \partial^i \mu+\dots
\end{equation}
where $\pi^i$ is the momentum density{, $\chi_{\pi\pi}$ the momentum static susceptibility},\footnote{In a Galilean invariant theory, this is simply $\chi_{\pi\pi}=m_{eff}n$ with $n$ the number density and $m_{eff}$ the effective mass, while in a Lorentz-invariant theory, this is $\chi_{\pi\pi}=\epsilon+p$ with $\epsilon$ the energy density and $p$ the pressure.} $\tau^{ij}$ the spatial stress tensor, and $\rho,s$ the charge and entropy densities. {The new transport coefficients $\lambda_{\rho,s,v}$ are defined more precisely in the following section, but can be thought of as out-of-equilibrium transport coefficients akin to the familiar shear and bulk viscosities. Physically, these transport coefficients cause a shift of the Drude weight in the thermoelectric conductivities. For instance, in the electric conductivity (see \eqref{acsigma} in the main text):
\begin{equation}
K_D=\frac{(\rho+\ell^2\lambda_\rho)^2}{\chi_{\pi\pi}+\ell^2\lambda_v}+\mathcal O(\ell^4)
\end{equation} 
Turning on a magnetic field, the cyclotron frequency, defined as the real part of the pole in the complex frequency plane, also receives $\mathcal O(\ell^2)$ contributions:
\begin{equation}
\omega_c= B\,\frac{\rho+2\ell^2\lambda_\rho}{\chi_{\pi\pi}+\ell^2 \l_v}+\mathcal O(\ell^4).
\end{equation}
}
{The Drude weight and the cyclotron mass can be measured experimentally (see eg \cite{Post:2020,Michon:2021,Legros:2022,VanHeumen:2022} for recent examples in cuprate high $T_c$ superconductors), and are often compared to thermodynamic data of the state, such as the number density and the effective mass. For instance, assuming Galilean invariance, $\lambda_\rho=0$ and 
\begin{equation}
K_D=\frac{ne^2}{m_{eff}}\left(1-\ell^2\frac{\lambda_v}{m_{eff}n}+\mathcal O(\ell^4)\right),\quad \omega_c=\frac{eB}{m_{eff}}\left(1-\ell^2\frac{\lambda_v}{m_{eff}n}+\mathcal O(\ell^4)\right).
\end{equation}
Our results show that in general the Drude spectral weight and the cyclotron frequency do not provide direct information on the the effective mass and the number density, as they receive corrections from explicit translation breaking. We comment on this further in the Discussion.}

Possible cyclotron frequency shifts due to interactions or impurities have been studied in the past. A result due to Kohn states that at zero temperature, the cyclotron frequency of a translation-invariant electron gas is insensitive to short-ranged interactions, \cite{PhysRev.123.1242}. However, this is no longer true in the presence of Umklapp or impurity scattering{, which cause a shift in the cyclotron frequency}, \cite{PhysRevB.31.3635,Kanki:1997} (see also the recent paper \cite{Guo:2023}). {Our work provides an effective field theory perspective on the effects of interaction and explicit breaking of translations on the cyclotron resonance. As is well-known, impurity scattering or absence of Galilean invariance also lead to a broadening of the cyclotron resonance, \cite{ando_theory_1975,Hartnoll:2007ih,mueller:2008}.}

We validate our analysis by matching these subleading terms to previously obtained results for strongly-coupled hydrodynamic metals modeled using gauge-gravity duality, \cite{Davison:2015bea,Blake:2015epa,Blake:2015hxa}.

We conclude with a discussion of implications of our results on (magneto-)transport experiments in strongly-coupled systems, and with an outlook on future research directions.

\section{Momentum-relaxed hydrodynamics with a scalar deformation}
We study the following system of dynamical equations:
\begin{equation}
\label{eomnoB}
\partial_t\epsilon+\partial_i \epsilon^i= - O_I \partial_t\Phi^I \,,\quad \partial_t\pi^i+\partial_j \tau^{ji} = O_I \partial^i\Phi^I \,,\quad \partial_t \rho+\partial_i j^i=0\,.
\end{equation}
They describe energy and momentum relaxation in a fluid by a set of scalar operators $O_I$, as well as the conservation of charge. $\epsilon$, $\pi^i$, $\rho$, $\epsilon^i$, $\tau^{ij}$, $j^i$, $O_I$ should be understood as the expectation values of the energy, momentum and charge densities and of the energy flux, spatial stress tensor, charge current and scalar operator in a finite density, thermal state.
The capital Latin indices $I, J \ldots$ run over all spatial dimensions $1\ldots d$, and while there is formally an internal symmetry rotating the scalars amongst themselves, it plays no role in this work, and we simply contract the $I, J$ indices with $\delta_{IJ}$. The $\Phi^I$ are the sources coupling to the $O_I$. Eventually, we will choose their background values to only depend on space, with a variation on large enough scales to be treated hydrodynamically, \cite{Lucas:2015lna}. It is useful to define the material derivative $\hat{D}\Phi_I \equiv (\partial_t+v^i\partial_i) \Phi_I$. We do not impose any specific boost symmetry at this stage, so we can more easily specialize to Galilean or Lorentz boost invariance later. However, we will not carry out a full-fledged analysis of boost-agnostic gradient corrections here, which is already rather intricate for momentum-conserving fluids, \cite{Novak:2019wqg,Armas:2020mpr}. We will use the simplifying assumption that our equilibrium state does not include a background fluid velocity, and will only write down gradient corrections linear in the velocity. In appendix \ref{app:relhydro}, we provide a more complete analysis specializing to systems invariant under Lorentz boosts.

With our choice of scaling scheme, the constitutive relations for the densities of conserved quantities and associated spatial fluxes begin at $\mathcal{O}(\ell^0)$,\footnote{We have implicitly made a frame choice for the hydrodynamic variables $T, \m, v^i$ such that the densities $\epsilon,\rho,\pi^i$ do not receive out of equilibrium gradient corrections, although, as we will shortly see, they do receive hydrostatic gradient corrections.}
\begin{equation}
\label{constrelgen}
\begin{split}
&\epsilon=\sum_{a=0}^\infty\epsilon_{(2a)}\,,\quad \rho=\sum_{a=0}^\infty\rho_{(2a)}\,,\quad \pi^i=\sum_{a=0}^\infty \pi^i_{(2a)}\, ,
\\
&\epsilon^i=\sum_{a=0}^\infty\epsilon^i_{(2a)}\,,\quad \tau^{ij}=\sum_{a=0}^\infty\tau^{ij}_{(2a)}\,,\quad j^i=\sum_{a=0}^\infty j^i_{(2a)}\,,\quad O^I=\sum_{a=0}^\infty O^I_{(2a+1)}\, ,
\end{split}
\end{equation}
where the subscript denotes the order in $\ell$. Inserting \eqref{constrelgen} into \eqref{eomnoB}, the dynamical equations can in turn be expanded order by order in $\ell^{2a+2}$, $a\geq0$, with ideal $a=0$ terms contributing at order $\ell^2$ and e.g.~viscous $a=2$ terms at order $\ell^4$. The scalar thermal expectation value starts at order $a=0$, i.e.~$\mathcal{O}(\ell)$, and thus contributes in \eqref{eomnoB} at the same order $\ell^2$ as ideal terms from the fluxes.

We now proceed to determine the divergence of the entropy current on-shell.
The first law of thermodynamics and the Smarr relation are given at ideal order by
\begin{equation}
d\epsilon_{(0)}=T ds_{(0)}+\mu d\rho_{(0)}+v_i d \pi^i_{(0)}\,,\quad \epsilon_{(0)}+p_{(0)}=Ts_{(0)}+\mu \rho_{(0)}+v_i\pi^i_{(0)}\,,
\end{equation}
where $s_{(0)}$ is the leading term in the expansion of the entropy density $s$ as in \eqref{constrelgen}.
The ideal constitutive relations\footnote{The $SO(d)$ spatial rotation invariance implies $\tau^{ij} = \tau^{ji}$, which leads to $\pi^i \propto v^i$.}
\begin{equation}
\label{constrelideal}
\epsilon^i_{(0)}=(\epsilon_{(0)}+p_{(0)})v^i\,,\quad \tau^{ij}_{(0)}=p_{(0)}\delta^{ij}+v^i\pi^j_{(0)}\,,\quad j^i_{(0)}=\rho_{(0)} v^i\,, \quad s^i_{(0)}=s_{(0)} v^i\,, 
\end{equation}
do not produce entropy, with $s^i_{(0)}$ the ideal entropy flux. Including gradient corrections, the divergence of the entropy current is on-shell
\begin{equation}
\label{entropydivergence}
T\partial_t s_{(0)}+T\partial_i s^i=-\sum_{a=1}^\infty \left( j^i_{(2a)}\partial_i\mu+ s^i_{(2a)}\partial_i T+\tau^{ij}_{(2a)}\partial_i v_j+O^I_{(2a+1)}\hat{D}\Phi_I\right)\equiv\Delta,
\end{equation}
with $s^i=\left(s_{(0)} v^i+\sum_{a=1}^\infty s^i_{(2a)}\right)$ and
\begin{equation}
\label{energyflux}
\epsilon^i_{(2a)}=T s^i_{(2a)}+\mu  j^i_{(2a)}+v_j\tau^{ij}_{(2a)}\,.
\end{equation}
Then, the second law of thermodynamics requires the derivative corrections to be such that $\Delta\geq0$ in \eqref{entropydivergence}. 

We are now in a position to discuss the gradient corrections appearing in the constitutive relations order by order in $\ell$, which are of three types: hydrostatic, non-hydrostatic dissipative, and non-hydrostatic non-dissipative. 

The hydrostatic terms do not contribute to the right-hand side of \eqref{entropydivergence}, originate from derivative corrections to the pressure functional in the static partition function and do not vanish on thermal equilibrium configurations, \cite{Armas:2020mpr}. When translations are not broken and at zero background fluid velocity, or in the presence of relativistic symmetry, these terms are absent to first order in gradients. In our case, two of them contribute to the pressure at order $a=1$, 
\begin{equation}
p_{(2),hs}=\Theta(\partial_i\Phi_I)^2+\tilde\Theta(v^i\partial_i\Phi_I)^2\, ,
\end{equation}
where $\Theta, \tilde{\Theta}$ are functions of the equilibrium scalars $T, \mu, v^2$, leading to an $a=1$ contribution to the first law of thermodynamics:
\begin{equation}
\label{hydrostatic1stlaw}
dp_{(2),hs}=s_{(2),hs}dT+\rho_{(2),hs} d\mu+\pi^i_{(2),hs}dv_i+2\Theta\partial^i\Phi^I d\partial_i\Phi_I+2\tilde\Theta v^i\partial_i\Phi^I v^j d\partial_j\Phi_I\,.
\end{equation}
The associated corrections to the entropy, charge, momentum and energy densities are
\begin{equation}
\label{hydrostatic1point}
\begin{split}
s_{(2),hs}=&\partial_T\Theta(\partial_i\Phi_I)^2+\partial_T\tilde\Theta(v^i\partial_i\Phi_I)^2\,,\quad
\rho_{(2),hs}=\partial_\mu\Theta(\partial_i\Phi_I)^2+\partial_\mu\tilde\Theta(v^i\partial_i \Phi_I)^2\,,\\
\pi^j_{(2),hs}=&(\partial\Theta/\partial v_j) (\partial_i\Phi_I)^2+(\partial\tilde\Theta/\partial v_j)(v^i\partial_i\Phi_I)^2 +2\tilde\Theta(v^i\partial_i\Phi_I)\partial^j\Phi^I\,, \\
\epsilon_{(2),hs} &=-p_{(2),hs}+Ts_{(2),hs}+\mu\rho_{(2),hs} +v_i\pi^i_{(2),hs} \,.
\end{split}
\end{equation}
Physically, these terms capture the deviations in thermodynamic quantities from homogeneity caused by the weak breaking of translations.

As these are hydrostatic corrections, they must not contribute to the right-hand side of \eqref{entropydivergence}, which requires the following $a=1$ corrections to the ideal constitutive relations of the energy flux, spatial stress tensor and to the scalar expectation value:\footnote{In deriving these expressions, we used that in thermal equilibrium, $\hat D\Phi_i=0$.}
\begin{align}
j^i_{(2),hs} &=\rho_{(2),hs}v^i\,,\quad s^i_{(2),hs} =s_{(2),hs} v^i\, ,\nonumber\\
\e^i_{(2),hs} &= \left(\epsilon_{(2),hs}+p_{(2),hs} \right)v^i - 2\Theta( v^j\partial_j \Phi_I )\del^i \Phi^I - 2\tilde{\Theta} ( v^j \del_j \Phi^I)^2v^i\, ,\nonumber\\
\tau^{ij}_{(2),hs}&=p_{(2),hs}\delta^{ij}-2\Theta\partial^i\Phi_I\partial^j\Phi^I+v^{(i} \left[(\partial\Theta/\partial v_{j)}) (\partial_k\Phi_I)^2+(\partial\tilde\Theta/\partial v_{j)})(v^k\partial_k\Phi_I)^2\right] , \label{hydrostatictauO}\\
 O^I_{(3),hs}&=-2\partial_i\left(\Theta\partial^i\Phi^I+\tilde\Theta v^i v^j\partial_j\Phi^I\right).\nonumber
\end{align}
We note that some of these hydrostatic corrections can be compactly included by replacing $(\epsilon_{(0)},p_{(0)},\rho_{(0)},s_{(0)},\pi^i_{(0)})\mapsto(\epsilon,p,\rho,s,\pi^i)$ in the densities and spatial fluxes. Those that cannot cancel out in the equations of motion, which can be seen using the $a=2$ first law \eqref{hydrostatic1stlaw}, and hence do not affect the retarded Green's functions of conserved operators.\footnote{They would affect the retarded Green's functions of non-conserved operators such as $\e^i, \tau^{ij}$ and $O^I$, but computing these requires coupling to Aristotelian background sources, \cite{Armas:2020mpr}, which is beyond the scope of this work.}
We will therefore not write hydrostatic corrections explicitly in what follows but it should be understood that all thermodynamic quantities from now on include such corrections. In appendix \ref{app:relhydro}, we show that \eqref{hydrostatic1stlaw} and \eqref{hydrostatictauO} follow from a hydrostatic partition function, by specializing to Lorentz-invariant theories and coupling to background sources.

The non-hydrostatic terms are proportional to combinations of the hydrodynamic fields that vanish on equilibrium configurations. Of these, the non-dissipative terms cancel out from the right-hand side of \eqref{entropydivergence}, do not produce entropy and obey equality-type relations. On the other hand, dissipative terms do not cancel out from the right-hand side of \eqref{entropydivergence}, and typically obey inequality-type relations to ensure that entropy only increases with time.

The non-hydrostatic terms in the constitutive relations \eqref{constrelgen} that contribute to order $\mathcal O(\ell^4)$ in the equations of motion \eqref{eomnoB} are (see also \cite{Armas:2021vku,Armas:2022vpf}):
\begin{subequations}
\label{AllConstRels}
\begin{align}
j^i_{(2),nhs}=&-\sigma_o \partial^i\mu-\alpha_o \partial^i T+\lambda_\rho \partial^i\Phi^I \hat{D}\Phi_I\, , \label{constrelj}\\
s^i_{(2),nhs}=&-\alpha_o \partial^i\mu-\f{\bar\kappa_o}{T} \partial^i T+\lambda_s \partial^i\Phi^I \hat{D}\Phi_I\, , \label{constrels}\\
\tau^{ij}_{(2),nhs}=&-\eta \left(\partial^i v^j+\partial^j v^i-\frac{2}d\delta^{ij}\partial\cdot v\right) - \frac{2\zeta}{d}\delta^{ij}\partial\cdot v \, , \label{constreltau}\\
O^I_{(1),nhs}=&-\gamma_1 \hat{D}\Phi^I\, , \label{constrelOphi1}\\
O^I_{(3),nhs}=&-\gamma_2 (\partial_i\Phi_J)^2 \hat{D}\Phi^I-\gamma_3\partial_i\Phi^I\partial^i\Phi^J \hat{D}\Phi_J-\lambda_\rho \partial^i\mu \partial_i\Phi^I-\lambda_s \partial^iT \partial_i\Phi^I - \l_v \del_t \hat{D}\Phi^I\, , \label{constrelOphi3}
\end{align}
\end{subequations}
where we have already imposed Onsager reciprocity under time reversal. Here $\s_o , \a_o$ and $\bar{\k}_o$ are the incoherent conductivities which do not contribute to the Drude peak in the ac thermoelectric conductivities, while $\eta, \z$ respectively denote the shear and bulk viscosity. At order $a=1$, we only wrote the subset of terms which survive when the background configuration of the $\Phi_I$'s only depends on space [e.g.~no $(\partial_t\Phi_J)^2 \hat D\Phi^I$ term in \eqref{constrelOphi3}].

At order $a=0$, $\gamma_1$ is the single derivative correction contributing to the dynamical equations, besides ideal terms, with the constraint $\gamma_1\geq0$ from the positivity of entropy production. {$\gamma_1$ captures the leading contribution to momentum relaxation, with \eqref{constrelOphi1} matching e.g.~(43) of \cite{Lucas:2015lna} after Fourier-transforming in space. Restricting to a single scalar $I=1$, the momentum relaxation rate of the system becomes
\begin{equation}
\label{MomRel}
\Gamma=\sum_{{\bf k}}k_i k_j|\Phi({\bf k})|^2\lim_{\omega\to0}\frac1\omega\,\textrm{Im}G^R_{O_1O_1}(\omega,{\bf k})\,.
\end{equation}
This is the same result as can be obtained using the memory matrix formalism, \cite{forster,Hartnoll:2008hs}.
If we assume that the chemical potential of the system is spatially-varying and is the source of the breaking of translations, then one can identify $\Phi(x)=\mu(x)$ and $O_1=\rho$. The momentum relaxation rate is then evaluated by computing the density-density retarded Green's function $G^R_{\rho\rho}(\omega,{\bf k})$ in the translation-invariant state. Doing so recovers the results of \cite{Andreev:2010,Lucas:2015lna,Li:2020}, e.g.~(48) of \cite{Lucas:2015lna}. Ref.~\cite{Chagnet:2023xsl} was also able to show that the memory matrix prediction and the momentum relaxation rate obtained by perturbing around a periodic chemical potential agree. This clarifies the physical meaning of the operators $O_I$: to leading order, they can be identified with the operators conjugate to spatially-dependent sources in the system. We have commented explicitly on the case of a spatially-dependent chemical potential, but it is straightforward to extend to a spatially-dependent temperature.}

At order $a=2$, positivity of entropy production yields the constraints
\begin{equation}
\sigma_o,\bar\kappa_o,\eta,\zeta\geq0\,,\quad T \alpha_o^2\leq\sigma_o\bar\kappa_o\,
\end{equation}
on the various non-hydrostatic dissipative transport coefficients.
The $\gamma_{2,3}$ and $\l_v$ do not get constrained at this order. $\lambda_{\rho,s}$ are of a different nature, since they drop out entirely from \eqref{entropydivergence}: they are non-dissipative, non-hydrostatic derivative corrections.\footnote{As we show in appendix \ref{app:relhydro} by restricting to the relativistic case, non-hydrostatic terms can be dissipative or non-dissipative, depending upon the choice of frame and the set of derivative terms.} In a previous version of this work, we did not write the $\lambda_v$ term, since for time-independent $\Phi_I$'s, it can be absorbed into a redefinition of $\lambda_{\rho,s}$ using the $a=0$ equations of motion. We have included it here in order to make our analysis more general. Keeping the coordinate dependence of the $\Phi_I$'s arbitrary for now also emphasizes that the $\lambda_{\rho,s}$ terms are not simply redefinitions of thermodynamics, since they involve the material derivative $\hat D\Phi_I$. 

For simplicity and with an eye towards widely used models of hydrodynamic metals in gauge/gravity duality, we now specialize to a linear profile for the translation-breaking sources $\Phi_I=\ell \delta_{Ii}x^i$ in \eqref{AllConstRels}, which preserves the isotropy of the state. This leads to the improved momentum-relaxation equation 
\begin{equation}
\label{MomRelaxEq2}
\left(1+\f{\ell^2 \l_v}{\chi_{\pi\pi}}\right)\left(\partial_t +\Gamma\right)\pi^i+\partial_j\tau^{ji}=-\ell^2\lambda_s \partial^i T-\ell^2\lambda_\rho \partial^i \mu+\mathcal O(\ell^6)\, ,
\end{equation}
with the momentum relaxation rate $\Gamma=\ell^2(\gamma_1+d\ell^2\gamma_2+\ell^2\gamma_3)/(\chi_{\pi\pi} + \ell^2 \l_v)$ receiving both $\mathcal{O}(\ell^2)$ and $\mathcal{O}(\ell^4)$ contributions. Also, $\chi_{\pi^i\pi^j}\equiv\delta\pi^i/\delta v_j$ denotes the momentum static susceptibility, 
\begin{equation}
\label{chipipi}
 \chi_{\pi^i\pi^j}=\frac{\partial^2p_{(0)}}{\partial v_i\partial v_j} +  d\ell^2\frac{\partial^2\Theta}{\partial v_i\partial v_j}+\ell^2 v^2 \frac{\del^2 \tilde{\Theta}}{\del v_i \del v_j}+4\ell^2v^{(i}\frac{\partial\tilde\Theta}{\partial v_{j)}} +2\ell^2\tilde\Theta\delta^{ij}\,.
\end{equation}
It is diagonal in the states with vanishing background fluid velocity we are interested in here, and so we have defined $\chi_{\pi^i\pi^j}=\chi_{\pi\pi}\delta^{ij}$. 
In \eqref{MomRelaxEq2}, all $\lambda_{\rho,s,v}$ terms come with a factor of $\ell^2$, and multiply derivatives of $\mu,\,T,\pi^i$, so that they indeed appear at subleading order compared to $\Gamma$, but at the same order as e.g.~viscous corrections to $\tau^{ij}$. 

We note that the extra terms on the right-hand side of \eqref{MomRelaxEq2} are natural to write down on general grounds at subleading order in the gradient expansion, irrespective of the specific mechanism responsible for breaking the spatial translation invariance. Thus we expect them to be present in linear response in other cases as well, such as for randomly-disordered sources. 

We now linearize about static equilibrium $V=\bar V+\delta V e^{-i\omega t+i k x}$, $\Sigma=\bar \Sigma+\delta \Sigma \, e^{-i\omega t+ik x}$, where $V=(\rho, s,\pi^i)$ is the vector of expectation values of the densities of conserved quantities (entropy is conserved at the linearized level), and $\Sigma =(\mu,T,v^i) $ is the vector of corresponding sources. We take $\bar v^i=\bar \pi^i=0$, with all other background values being non-vanishing constants without any spatial dependence. 

The retarded Green's functions in Fourier space are found using the standard result \cite{Kovtun:2012rj}
\begin{equation}
G^R(\omega,k)=M(k)\cdot\left(i\omega\chi-M(k)\right)^{-1}\cdot\chi\, ,
\end{equation}
where $\chi$ is the matrix of static susceptibilities and $M(k)$ is defined from the Fourier transform of the linearized dynamical equations $-i\omega \delta V+M(k)\cdot \delta\Sigma=0$. Taking the $k\to0$ limit and using the linearized charge and entropy conservation equations, we obtain the ac electric, thermoelectric and thermal conductivities, $\sigma(\omega)$, $\alpha(\omega)$ and $\bar{\kappa}(\omega)$ respectively, as
\begin{align}
\label{acsigma}\sigma(\omega)\equiv \frac{i}\omega \, G^R_{\vec J\vec J}(\omega,k=0)&=\sigma_o+\frac{(\rho+\ell^2\lambda_\rho)^2}{\chi_{\pi\pi}+\ell^2 \l_v}\,\frac1{\Gamma-i\omega}+\mathcal O(\ell^2)\, ,\\
\label{acalpha}\alpha(\omega)\equiv \frac{i}\omega \, G^R_{\vec J\vec S}(\omega,k=0)&=\alpha_o+\frac{(\rho+\ell^2\lambda_\rho)(s+\ell^2\lambda_s)}{\chi_{\pi\pi}+\ell^2 \l_v}\,\frac1{\Gamma-i\omega}\, +\mathcal O(\ell^2),\\
\label{ackappa}\bar\kappa(\omega)\equiv \frac{iT}\omega \, G^R_{\vec S\vec S}(\omega,k=0)&=\bar\kappa_o+T\frac{(s+\ell^2\lambda_s)^2}{\chi_{\pi\pi}+\ell^2 \l_v}\,\frac1{\Gamma-i\omega}+\mathcal O(\ell^2)\, ,
\end{align}
where $\vec J$, $\vec S$ are the charge and entropy spatial currents. In these expressions, it should be understood that the numerator is known to $\mathcal O(\ell^2)$, while the denominator is known to $\mathcal O(\ell^4)$.

As expected, the expressions above feature a Drude-like pole at $\omega=-i\Gamma$, as well as incoherent (analytic) contributions $\sigma_o$, $\alpha_o$ and $\bar\kappa_o$. These represent relaxation processes which do not drag momentum, \cite{Davison:2015taa,Davison:2015bea}, and thus are not sensitive to the Drude-like pole. They originate from the gradients of $\mu$ and $T$ corrections to the charge and heat currents \eqref{constrelj}, \eqref{constrels}. The Drude pole gives an $\mathcal{O}(\ell^{-2})$ contribution to the dc conductivity, obtained by taking the $\omega\to0$ limit. Incoherent contributions appear at order $\mathcal{O}(\ell^0)$. The non-dissipative, non-hydrostatic terms $\lambda_{\rho,s}$, along with $\l_v$, renormalize the Drude weight $K_D$, which is no longer simply given by thermodynamic data of the state. E.g.~for the electric conductivity,
\be
\label{DrudeWeight}
K_D=\frac{(\rho+\ell^2\lambda_\rho)^2}{\chi_{\pi\pi} + \ell^2 \l_v}\,.
\ee
The $\l_{\rho,s}$ should not be confused with gradient corrections in the hydrostatic partition function \cite{Banerjee:2012iz,Jensen:2012jh,Armas:2020mpr}, which are also non-dissipative but only act to change the expectation values of the densities of conserved operators away from their values in the translation invariant theory and which have already been accounted for in $\rho$ and $s$ in \eqref{acsigma}-\eqref{ackappa}, \emph{cf.}~\eqref{hydrostatic1stlaw}. In the $\omega\to0$ limit, $\lambda_{\rho,s,v}$ contribute to the dc conductivities at the same order $\mathcal{O}(\ell^0)$ as the dissipative, incoherent terms $\sigma_o$, $\alpha_o$ and $\bar{\kappa}_o$. 

The collective modes, which are obtained by looking for poles of the retarded Green's functions, have the same general structure as known from earlier literature, \cite{Hartnoll:2007ih}. In the transverse sector, there is a gapped mode capturing the relaxation of transverse momentum:
\begin{equation}
\label{transversediff}
\omega_\perp = -i\Gamma-i\frac{\eta}{\chi_{\pi\pi}}k^2+\dots 
\end{equation}
In the longitudinal sector, there is an  $\mathcal O(\ell^4)$ mode quadratic in $k$ representing thermal diffusion, with no $\mathcal{O}(\ell^2)$ corrections to the diffusivity at this order:
\begin{equation}
\label{longdiff}
\omega_D=-i D_o k^2+\mathcal O(\ell^6)\,,\quad D_o  = \frac{ s^2\sigma_o+\f{\rho^2}{T} \bar{\kappa}_o-2\rho s\alpha_o}{ s^2\chi_{\rho\rho}+\rho^2\chi_{ss}-2 s\rho\chi_{\rho s}}\, ,
\end{equation}
and two $\mathcal O(\ell^4)$ modes representing pressure and momentum fluctuations crossing over between overdamped and underdamped behavior as a function of $k$ at fixed $\ell$:
\begin{equation}
\label{long}
\omega^\pm=-\frac{i}{2}\Gamma\pm\frac12\sqrt{4 v_s^2k^2-\Gamma^2}+\mathcal{O}(ik^2)+\mathcal{O}(k \ell^2)+\dots\,,\quad v_s^2=\frac{\rho^2\chi_{ss}-2s\rho\chi_{\rho s}+s^2\chi_{\rho\rho}}{\chi_{\pi\pi}(\chi_{\rho\rho}\chi_{ss}-\chi_{\rho s}^2)}\, ,
\end{equation}
where $v_s$ is the speed of sound in the theory invariant under translations, given by various static susceptibilities of charge and entropy. At the order at which we are working, there are also $\mathcal{O}(i k^2,k\ell^2)\sim \mathcal{O}(\ell^4)$ corrections that our theory captures (going beyond expressions previously reported, eg in \cite{lucas_sound_2016}, which only worked to leading order in disorder perturbation), but which have expressions complicated enough that we do not report them here explicitly. {The effects of the $\lambda_{\rho,s,v}$ terms are essentially to renormalize the sound velocities, $v_s^2\mapsto v_s^2+\#\ell^2$. Note that in \eqref{long}, $\rho$ and $s$ do include $\Theta$ and $\tilde\Theta$ corrections as well.}

While it seems possible from the right-hand side of \eqref{MomRelaxEq2} to reabsorb the $\lambda_{\rho,s}$ terms into a redefinition of pressure in $\tau_{ij}$, and from the ac conductivities into a redefinition of $\rho$ and $s$, the thermodynamic terms appearing in retarded Green's functions at finite $k$ are not compatible with such a redefinition. Also one should recall that these terms really come from the material derivative $\hat D\Phi_I$, which is intrinsically non-hydrostatic.

In experiments, it is the thermal conductivity with open circuit boundary conditions $\vec J=0$, given by $\kappa\equiv\bar\kappa-T\alpha\cdot\sigma^{-1}\cdot\alpha$, which is measured rather than thermal conductivity at zero electric field $\bar{\k}$. To the order we are working, we find
\begin{equation}
\kappa=\bar\kappa_o+\frac{Ts^2}{\rho^2}\sigma_o-2\frac{Ts}\rho\alpha_o+\mathcal O(\ell^2)\,.
\end{equation}
As is well-known, the Drude-like peak drops out from this observable, while the Wiedemann-Franz law which holds in Fermi liquids is violated in the hydrodynamic regime, \cite{Mahajan:2013cja,Crossno:2015}. 

It is worthwhile commenting upon various limiting cases. In the Galilean limit, the Galilean boost Ward identity implies $\pi^i=m_{\rm eff}\, j^i/e$, where $m_{\rm eff}$ is the effective mass and $e$ the unit electric charge. Further, the pressure and the coefficients of the derivative terms become functions of $T$ and $\mu_{\rm gal}=\mu+m_{\rm eff}v^2/2$, the Galilean boost-invariant definition of the chemical potential, see e.g.~\cite{Armas:2020mpr}. Defining $\rho=n e$, with $n$ being the number density, we have
\begin{equation}
\label{Galilean}
 \chi_{\pi\pi}= n\, m_{\rm eff}\,,\quad \sigma_o=\alpha_o=\l_\rho=0\,,\quad \tilde\Theta=0\,.
\end{equation}
The last constraint comes from imposing the Ward identity on the hydrostatic sector.

On the other hand, the relativistic limit implies $\epsilon^i=\pi^i$. Further, the pressure and the coefficients of the derivative terms become functions of $T_{\rm rel}=T/\sqrt{1-v^2}$ and $\mu_{\rm rel}=\mu/\sqrt{1-v^2}$,\footnote{We work with natural units and have set the speed of light to unity.} which are the Lorentz boost-invariant definitions of the temperature and chemical potential, see e.g.~\cite{Armas:2020mpr} for further details. Then one has
\begin{equation}
\label{Lorentz}
\chi_{\pi\pi}=Ts+\mu \rho=T_{\rm rel}s_{\rm rel}+\mu_{\rm rel}\rho_{\rm rel}\,,\quad \frac{\bar\kappa_o}T=-\frac{\mu}T\alpha_o=\frac{\mu^2}{T^2}\sigma_o\,, \quad  \l_s=-\frac{\m}T\, \l_\rho\,,\quad \tilde\Theta=-\Theta.
\end{equation}
Here $s_{\rm rel}=s\sqrt{1-v^2}$ and $\rho_{\rm rel}=\rho \sqrt{1-v^2}$. The constraint $\tilde \Theta=-\Theta$ agrees with our relativistic analysis in Appendix \ref{app:hydrostatic}.\footnote{A complete match with Appendix \ref{app:relhydro} would require to take into account all derivative corrections non-linear in the fluid velocity, which we neglected in \eqref{constrelj}-\eqref{constrelOphi3}, and mapping our choice of frame to the Landau frame. For our purposes, it is enough to impose \eqref{Lorentz}.}

These expressions allow us to reproduce the results obtained in \cite{Davison:2015bea,Blake:2015epa,Blake:2015hxa}, modeling strongly-coupled metallic phases using the gauge/gravity duality. Specifically, in the model \cite{Andrade:2013gsa}
\begin{equation}
S=\int d^2x dt dr \sqrt{-g}\left[R-\frac12(\nabla\Phi_I)^2-\frac14 F^2-2\Lambda\right],
\end{equation}
with $\Phi_I=\bar\Phi_I$,
the conductivities take the form \eqref{acsigma}-\eqref{ackappa} subject to Lorentz symmetry \eqref{Lorentz}, \cite{Davison:2015bea,Blake:2015hxa}, where (dropping $rel$ subscripts)
\begin{equation}
\label{matchholo}
\lambda_s=-\frac{\ell^2n^2s}{\epsilon+p}\,,\quad \lambda_v=-\frac{\ell^2n^2}{\epsilon+p}-\ell^2(\epsilon+p)\l(T,\mu)\,,\quad \Gamma=\frac{s\ell^2}{4\pi(\epsilon+p)}\left(1+\ell^2\,\lambda(T,\mu)\right),
\end{equation}
and the function $\lambda(T,\mu)$ was computed for these holographic states in  \cite{Davison:2015bea,Blake:2015hxa}. We recall its expression in appendix \ref{app:comp_holo}, together with the expressions for the thermodynamic quantities, while $\Theta$ is given in \eqref{Thetaholstate}.

In the holographic examples it was assumed that $\ell\ll\mu, T$, meaning that the translation symmetry-breaking scale is smaller than both temperature and chemical potential, and temperature was the largest scale in the system. Then, the background geometry captures small $T$ and $\mu$ perturbations away from asymptotically Anti-de Sitter spacetimes. However, the effective theory just developed also applies in the vicinity of zero temperature critical phases with an emergent translation symmetry, \cite{Davison:2018nxm}, i.e.~in the low temperature regime $\ell,T\ll\mu$. We leave exploring this for future work.

\section{Magnetotransport}
We now specialize to two spatial dimensions, $d=2$, and include an external (non-dynamical) magnetic field through $F^{ij}=B\varepsilon^{ij}$, with $\varepsilon^{ij}$ being the anti-symmetric Levi-Civita tensor. Along the lines of \cite{Hartnoll:2007ih}, we adopt the scaling $B\sim\ell^2$, which means the magnetic field contributes to the constitutive relations at the same order as the derivatives of $\mu$, $T$ and $v^i$. It appears as an extra term on the right-hand side of the momentum equation in \eqref{eomnoB}, which now becomes
\begin{equation}
\partial_t\pi^i+\partial_j\tau^{ji} = O_I  \partial^i\Phi^I + F^{ik} j_k\,.
\end{equation}
This term is responsible for cyclotron-like orbital motion, and generates the cyclotron frequency appearing as real part of the spectrum of hydrodynamic poles, $\omega=\pm( \rho B/\chi_{\pi\pi})+\ldots$. 
Evaluating the divergence of the entropy current results in the same ideal constitutive relations for the spatial fluxes, while simply affecting the right-hand side of \eqref{entropydivergence} through the replacement $\partial^i\mu\mapsto\partial^i\mu-F^{ik}v_k$. Apart from the same shift in $\partial^i\mu$, no new $\mathcal{O}(\ell^2)$ terms (hydrostatic or non-hydrostatic) are generated in the constitutive relations for the spatial fluxes \eqref{constrelj}-\eqref{constreltau} or in the scalar expectation value \eqref{constrelOphi1}, \eqref{constrelOphi3}.\footnote{``Hall''-like terms such as $B \varepsilon^{ij}\partial_j\mu$ would appear at $\mathcal{O}(\ell^4)$ in the constitutive relations.} Evaluating them for $\Phi_I=\ell \delta_{Ii} x^i$, the momentum relaxation equation becomes
\begin{equation}
\label{MomRelaxEqB}
\left(1+\f{\ell^2\l_v}{\chi_{\pi\pi}}\right)\left(\partial_t \pi^i+ \G_c \pi^i-\omega_c\varepsilon^{ij}\pi_j\right)+\partial_j\tau^{ji} = -\ell^2\lambda_s \partial^i T-\ell^2\lambda_\rho \partial^i \mu-B\sigma_o\varepsilon^{ij}\partial_j\mu-B\alpha_o\varepsilon^{ij}\partial_jT\, ,
\end{equation}
where 
\begin{equation}
\omega_c\equiv B\,\frac{\rho+2\ell^2\lambda_\rho}{\chi_{\pi\pi}+\ell^2 \l_v}\,,\qquad \Gamma_c\equiv\Gamma+\frac{\s_o B^2}{\chi_{\pi\pi}}\, ,
\end{equation}
with $\G$ defined below \eqref{MomRelaxEq2}. At zero wave-vector, collective excitations go around damped cyclotron orbits, i.e., the retarded Green's functions have poles at 
\begin{equation}
\omega=\pm\omega_c-i\Gamma_c+\dots \, .
\end{equation} 
We observe that the cyclotron frequency also receives $\mathcal{O}(\ell^2)$ corrections compared to its translation-invariant expression. {Such corrections have not appeared in previous literature (e.g.~\cite{Hartnoll:2007ih,mueller:2008}), to the best of our knowledge. To better understand their effect, it is helpful to consider the Galilean and the Lorentzian limits, \eqref{Galilean} and \eqref{Lorentz}. The Galilean limit is of most relevance to eg overdoped cuprate high $T_c$ superconductors, as we comment upon in the Discussion, while the Lorentz limit is useful for Graphene near the charge neutrality point, where an approximate hydrodynamic regime emerges,\cite{Lucas:2017idv}. In the Galilean limit, the shift of the cyclotron frequency reads}
\begin{equation} 
\omega_c^{gal}=\frac{eB}{m_{eff}}\left(1-\ell^2\frac{\lambda_v}{m_{eff}n}\right).
\end{equation}
{Thus the shift in the cyclotron frequency from its standard expression $eB/m_{eff}$ can be used to measure the coefficient $\lambda_v$. $m_{eff}$ is the effective mass of the charge carriers and can be determined by other means, eg by specific heat measurements of quantum oscillations. As we comment upon in the Discussion, another way to determine $\lambda_v$ is through a determination of the spectral weight in the Drude peak observed in the ac conductivity. In the Galilean limit, the Drude spectral weight \eqref{DrudeWeight} becomes }
\begin{equation}
K_D=\frac{n e^2}{m_{eff}}\left(1-\ell^2\frac{\lambda_v}{m_{eff}n}\right).
\end{equation}
{On the other hand, the broadening $\Gamma_c$ of the cyclotron resonance, either due to the breaking of spatial translations or in non-Galilean invariant systems, is well-known. To the order in perturbation theory we are working, there are no new contributions to $\Gamma_c$ compared to earlier work.}

The thermoelectric conductivities are easier to present in complexified notation, e.g.~$\sigma_c\equiv\sigma_{xx}+i\sigma_{xy}$:
\begin{align}
\label{acsigmaB}\sigma_c(\omega)&=\sigma_o + \frac{(\rho+\ell^2\lambda_\rho+i B\sigma_o)^2}{(\chi_{\pi\pi} + \ell^2\l_v)(\G_c-i(\o + \o_c))}+\mathcal O(\ell^2)\, , \\
\label{acalphaB}\alpha_c(\omega)&=\alpha_o + \frac{(\rho+\ell^2\lambda_\rho+iB\sigma_o)(s+\ell^2\lambda_s+iB\alpha_o)}{(\chi_{\pi\pi} + \ell^2\l_v)(\G_c-i(\o + \o_c))}+\mathcal O(\ell^2)\, ,\\
\label{ackappaB}\bar\kappa_c(\omega)&=\bar\kappa_o + T \, \frac{(s+\ell^2\lambda_s+iB\alpha_o)^2}{(\chi_{\pi\pi} + \ell^2\l_v)(\G_c-i(\o + \o_c))}+\mathcal O(\ell^2)\, .
\end{align}

Specializing to phases with Lorentz boost invariance \eqref{Lorentz}, our results match the holographic results of \cite{Blake:2015hxa}.

There are various transport coefficients that are interesting to evaluate, with the ac conductivities at hand.
The Hall coefficient is defined as $R_H\equiv(\sigma^{-1})_{xy}/B$. We find
\begin{equation}
\label{HallCoeff}
R_H=-\frac1{\rho}+\mathcal O(\ell^4)\,,
\end{equation}
meaning that the Hall coefficient continues to be given by the charge density to the order we are working,as in the translation-invariant case.\footnote{Since dc conductivities are known to order $\mathcal O(\ell^2)$, resistivities are know to order $\mathcal O(\ell^4)$.} Note though that, as we have explained above, the charge density may receive hydrostatic gradient corrections capturing slow spatial variations in the local thermal equilibrium state. 
On the other hand, the thermal Hall conductivity with open circuit boundary conditions, $\kappa_\perp\equiv(\bar\kappa-T\alpha\cdot\sigma^{-1}\cdot\alpha)_{xy}$, vanishes to the order we are working,
\begin{equation}
\kappa_\perp = 0+\mathcal O(\ell^2)\, ,
\end{equation}
and so does the magnetoresistance $\rho_{\rm dc}/[\rho_{\rm dc}(B=0)]-1$.

\section{Discussion}
\label{sec:discussion}
In this work, we have constructed a consistent theory of hydrodynamic metals, which goes beyond the usual leading order momentum-relaxing contributions and includes new terms that formally appear at the same order as more familiar derivative corrections to the constitutive relations of spatial fluxes, such as the shear and bulk viscosities or the incoherent conductivities. Some of these terms are hydrostatic, meaning that their role is to model weak spatial inhomogeneities in the equilibrium states. There are also non-hydrostatic terms, which are intrinsically out-of-equilibrium derivative corrections

In linear response, these terms only contribute to the real part of collective excitations, for instance by shifting the speed of sound or the cyclotron frequency from their translation-invariant expressions. They also renormalize the Drude-like weights which appear in the frequency-dependent thermoelectric conductivities. 

An interesting transport coefficient is the Hall coefficient $R_H$. In Galilean, translation-invariant hydrodynamics at finite temperature, as well as in a Fermi liquid with a spherical hole-like Fermi surface near zero temperature, it evaluates to $R_H=1/(n_ee)$, where the density of electrons $n_e$ is defined from $\rho=-n_e e$. The Hall number $n_H\equiv 1/(R_H e) $ is consequently often used as a measure of the density of conduction electrons in metals, by turning on a sufficiently large magnetic field such that zero temperature is well-approximated.  

Away from Galilean boost and translation invariance, it is less clear that this result should continue to hold. Anisotropic Fermi surfaces \cite{Ong:2011} or the presence of Van Hove singularities \cite{Maharaj:2016} lead to a mismatch between the carrier density and the Hall number. How to reformulate such results in a hydrodynamic language is not known. The Hall coefficient has been computed in gauge/gravity duality non-perturbatively in both the magnetic field and strength of disorder \cite{Blake:2014yla}, generally showing deviations from $1/\rho$ when either becomes sizable. However, the dc result itself is not enough to elucidate whether such deviations are caused by non-hydrodynamic excitations or by higher-order gradient corrections. Experimentally, at large enough magnetic fields and low temperatures, $n_H$ in very overdoped cuprate high $T_c$ superconductors reaches a plateau, and appears to tend to the expected number density for a large hole-like Fermi surface, $n_H\simeq1+p$, \cite{badoux2016change,putzke2021reduced}. This happens in holography as well.

At more elevated temperatures ($15-40K$) and lower magnetic fields ($5-35T$), ref.~\cite{Legros:2022} measured the  cyclotron mass $m_c\equiv e B/\omega_c$ in the cuprate LSCO using optical spectroscopy to determine the cyclotron frequency. The cyclotron mass thus measured smoothly increases all the way to the highest doping measured, at the edge of the superconducting dome. This is in sharp contrast with the mass measured from the specific heat, which peaks at the doping where the pseudogap terminates and otherwise decreases towards larger doping, \cite{Michon2019,PhysRevB.103.214506}. This feature has been interpreted as the sign of a smeared out phase transition. In the Galilean limit, we find for the cyclotron mass $m_c=m_{eff}+\ell^2\lambda_v/n$. Recall that the sign of $\lambda_v$ is not constrained by entropy production and could be negative in principle. As doping is increased in the sample, so does disorder. This suggests $m_c$ should increase as well with doping, which is what \cite{Legros:2022} observes. Instead, the specific heat is sensitive to the hydrostatic term $\Theta$, which could cause a different dependence on doping, as observed in \cite{Michon2019,PhysRevB.103.214506}. Further, as $\Theta$ is a hydrostatic coefficient (i.e.~it contributes to derivatives of the free energy), its dependence on thermodynamic parameters reflects phase transitions, but not in the case of $\lambda_v$ as it is non-hydrostatic.
 
From $m_c$, \cite{Legros:2022} then deduced the cyclotron density $n_c$ by combining it with the optical spectral weight $K_D$. Intriguingly, they find rough agreement, $n_c\simeq n_H$, even though these measurements are done in different temperature and magnetic field regimes, which are closer to the regime our results apply to. Independent from whether the system has Galilean invariance or not, we find from \eqref{HallCoeff} that $n_H=-\rho+\mathcal{O}(\ell^4)$, i.e.~the leading $\mathcal{O}(\ell^2)$ corrections due to e.g.~$\lambda_{\rho,s,v}$ or $\sigma_o$ cancel out from the Hall coefficient, compatible with the measurements of \cite{Legros:2022}. It would be interesting to understand whether this property continues to hold at higher order in the gradient expansion, and if so, what protects it. In other words, the ratio of the Drude weight $K_D$, the cyclotron frequency and the Hall coefficient, $K_D R_H/(B\omega_c)$, appears to be less sensitive to the strength of translation-symmetry breaking than naively expected.

Transport experiments in overdoped cuprates have been argued as providing evidence that there is a coherent and an incoherent charge conduction channel, \cite{ayres2021incoherent,phillips2022stranger}. The incoherent channel can only arise if the theory is not invariant under Galilean boosts (which is perhaps not unreasonable to assume in cuprates given that their properties differ from those of a Fermi liquid), and in our hydrodynamic approach this is manifested by a nonzero $\sigma_0$. On the other hand, a thorough spectroscopic analysis \cite{VanHeumen:2022} suggests that while there is an incoherent contribution to charge transport in cuprates, it does not carry any spectral weight at zero frequency.  The  $\lambda_{\rho,s,v}$ terms that renormalize the Drude weight turn out to contribute to dc conductivities at the same order as the incoherent conductivity $\sigma_0$. Our results underline the need to combine transport and optic experiments in order to assess whether there is an incoherent sector and how it contributes to dc transport.

\section{Outlook}

There are a number of directions for future work. The $\lambda_{\rho,s,v}$ terms may affect the velocity profiles of hydrodynamic electronic flows \cite{fritz2023hydrodynamic} along the lines of e.g.~\cite{Levitov_2016}, as they appear at the same order in the hydrodynamic expansion as the shear viscosity. It would also be worth computing how these corrections arise from the Boltzmann equation in kinetic theory, \cite{fritz2023hydrodynamic,Lucas:2017vlc}, which is often used to interpret transport experiments in metals close to a Fermi liquid regime.

Our derivation of gradient corrections does not make any particular assumption on the profile of the scalar source breaking spatial translations, other than it should be slowly-varying compared to the local equilibriation scales that make the description of the system in terms of coarse-grained collective variables well-defined. On the other hand, we made a specific choice of linearly-dependent scalars to facilitate the analysis of linear response. Considering random spatial dependence along the lines of \cite{Lucas:2015lna,Scopelliti:2017sga} is a natural next step in order to address the physics of disorder, which is an interesting topic in itself, and further has recently been argued to play a defining role in strange metallic transport, \cite{Patel:2022gdh}. This will also allow to connect with the corresponding gauge-gravity calculations, \cite{Lucas:2014zea,Hartnoll:2014cua,Hartnoll:2015faa,Hartnoll:2015rza,Ganesan:2020wzm,Ganesan:2021gun,Huang:2023ihu}. 

Here we chose to introduce spatial dependence through a scalar operator, but it is also possible to do so through a spatially-varying chemical potential, \cite{Andreev:2010,Lucas:2015lna,Li:2020,Chagnet:2023xsl}. Then the constitutive relations are unchanged from the translation-invariant case. In this case, the operator $O$ on the right-hand side of the momentum equation should be identified with the inhomogeneous part of the charge density. The transport coefficients in the effective equation \eqref{MomRelaxEq} we arrived at will now be determined by the transport coefficients of the homogeneous state, such as the thermal diffusivity or shear and bulk viscosities, as is known to happen for the momentum relaxation rate \cite{Andreev:2010,Davison:2013txa,Lucas:2015lna,Li:2020,Chagnet:2023xsl}. Further, these results can be recovered by means of the memory matrix approach, \cite{forster}, see \cite{Davison:2013txa} for an illustration where $O$ is instead taken to be the energy density. We expect that a similar approach will yield explicit results for the transport coefficients $\lambda_{\rho,s,v}$ most of our analysis has focused on, although this may require to extend the memory matrix to subleading order in the disorder strength (see \cite{Lucas:2015lna} for some comments on this). Our results then provide a framework to interpret linear response in a more transparent manner.

Finally, on general grounds we anticipate the terms discussed here to arise in other hydrodynamic theories with approximate symmetries, such as pinned charge density waves \cite{Baggioli:2022pyb}, superfluids with vortices or more generally phases with broken symmetries and defects, \cite{Davison:2016hno,Delacretaz:2017zxd,Armas:2022vpf,Armas:2023tyx}, M\"uller-Israel-Stewart theories \cite{Grozdanov:2018fic} and holographic phases of matter, \cite{Karch:2009zz,Davison:2011ek,Chen:2017dsy,Grozdanov:2018fic,Davison:2018ofp,Davison:2022vqh}. Understanding both their structure \cite{Armas:2021vku,Armas:2023tyx} and their physical consequences will pave the way to their experimental determination, similar to experimental measurement of more familiar transport coefficients such as the shear viscosity.

\acknowledgments
We are grateful to Daniel Brattan, Nicolas Chagnet, Erik van Heumen, Steve Kivelson, Pavel Kovtun, Alex Levchenko, Dmitrii Maslov, Ben Withers and Vaios Ziogas for discussions.  We would like to thank Mike Blake, Richard Davison, Andy Lucas and especially Akash Jain and Eric Mefford for helpful feedback on a draft of this paper. The work of BG and AS is supported by the European Research Council (ERC) under the European Union's Horizon 2020 research and innovation programme (grant agreement No.~758759). BG and AS would like to acknowledge the hospitality received from the Nordic Institute for Theoretical Physics (Nordita), Stockholm, over the course of the program ``Recent developments in strongly correlated quantum matter,'' during which parts of this work were done. Part of this work was performed at the Aspen Center for Physics, which is supported by National Science Foundation grant PHY-2210452. This research was supported in part by the National Science Foundation under Grant No.~NSF PHY-1748958. AS would also like to acknowledge the hospitality of the Korea Institute for Advanced Study, Seoul, during the later stages of this project, as well as Nordita, Stockholm, during the workshop ``Hydrodynamics at all scales.''


\appendix
\section{Relativistic formulation\label{app:relhydro}}
In this appendix, we first provide details of the hydrostatic corrections by specializing to the relativistic case. Subsequently, working in the Landau frame, we discuss the out of equilibrium constitutive relations and constraints from positivity of entropy production.

\subsection{Hydrostatic derivative corrections}
\label{app:hydrostatic}
We will discuss the hydrostatic derivative corrections using the framework of the \emph{equilibrium generating functional}. We will focus on the relativistic i.e.~Lorentz-invariant case for the present analysis \cite{Banerjee:2012iz, Jensen:2012jh}. The hydrostatic corrections for the Galilean-invariant case will also have a structure similar to what we find below. Hydrostatic corrections for the boost-agnostic case can be studied along the lines of \cite{Armas:2020mpr}, but as we do not allow for a background fluid velocity, many of the inherent subtleties do not matter for the present discussion.

For a relativistic fluid with a global $U(1)$ symmetry, the state of thermal equilibrium in the presence of sources (i.e., the background metric $g_{\mu\nu}$ which sources the fluid stress tensor and the background gauge field $A_\mu$ that sources the conserved $U(1)$ current) implies the existence of a timelike Killing vector $V^\mu$ and a gauge scalar field $\Lambda_V$, such that
\begin{equation}
\pounds_V g_{\mu\n} = 0\, , \quad \pounds_V A_\mu + \del_\mu \L_V = 0.
\label{eq:Lie_dervs}
\end{equation}
Here $\pounds_V$ denotes the Lie derivative with respect to $V^\mu$. Under a gauge transformation of the background field $A_\mu \rightarrow A_\mu + \del_\mu \l$, the scalar field $\Lambda_V$ transforms as $\Lambda_V \rightarrow \Lambda_V - V^\mu \del_\mu \l$. This ensures that the equilibrium conditions in \eqref{eq:Lie_dervs} are gauge invariant.

The timelike Killing vector $V^\mu$ and the gauge scalar field $\L_V$ provide us with a natural definition for the temperature, chemical potential and fluid velocity, given by
\begin{equation}
T = \frac{T_0}{\sqrt{-V^2}}\, , \quad \mu = \frac{V^\mu A_\mu + \L_V}{\sqrt{-V^2}}\, , \quad u^\mu = \frac{V^\mu}{\sqrt{-V^2}}.
\label{eq:thermo_frame}
\end{equation}
Here $T_0$ is a normalization constant, and $V^2 \equiv g_{\mu\n} V^\mu V^\n$ is negative since $V^\mu$ is timelike. The definitions of $T, \mu$ and $u^\mu$ above are gauge invariant. Further, $T, \mu$ are invariant whereas $u^\mu$ is covariant under diffeomorphisms. The particular choices in \eqref{eq:thermo_frame} for the hydrodynamic variables define the \emph{thermodynamic frame}. Using these definitions, it is straight forward to check that
\begin{subequations}
\begin{align}
\pounds_V T &= \frac{T}{2} u^\alpha u^\beta \pounds_V g_{\alpha\beta}\, ,\\
\pounds_V \mu &=  \frac{\mu}{2} u^\alpha u^\beta \pounds_V g_{\alpha\beta} + u^\mu \left(\pounds_V A_\mu + \del_\mu \L_V\right),\\
\pounds_V u^\mu &= \frac{1}{2} u^\mu u^\alpha u^\beta \pounds_V g_{\alpha\beta}\, .
\end{align}
\end{subequations}
By virtue of \eqref{eq:Lie_dervs}, one has $\pounds_V T = \pounds_V \mu = \pounds_V u^\mu = 0$, as expected in thermal equilibrium, where all quantities should be independent of time. 

In addition, to model the weak explicit breaking of spatial translation invariance, we also have a set of scalar fields $\P^I$. These must also satisfy $\pounds_V \Phi^I = 0$ in thermal equilibrium. 

With these inputs, and the scaling $\del T \sim \del \mu \sim (\del\P)^2 \sim \ell^2$, the leading terms in the equilibrium generating functional take the form \cite{Banerjee:2012iz, Jensen:2012jh}
\begin{equation}
\mathbb{W}[g_{\mu\n}, A_\mu, \P^I] = \int d^{d+1}x \sqrt{-g} \left[\bar{p}(T,\mu) + \Theta(T,\mu)\, (\partial\Phi)^2 \right] ,
\label{eq:gen_func}
\end{equation}
where ${\bar p}(T,\mu)$ denotes the fluid pressure without any translation breaking sources turned on,\footnote{If the $\Phi^I$ have a homogeneous background value, then $\bar p$ depends on this quantity as well.} while the corrected pressure is $p = {\bar p} + \Theta \, (\partial\Phi)^2$ with $(\partial\Phi)^2=\del_\a \P_I \, \del^\a \P^I$. Since we are not considering nonzero background fluid velocities, $\Theta(\del\P)^2$ is the only hydrostatic correction entering at this derivative order (the other linearly independent term we could write at this order is proportional to $(u^\mu\nabla_\mu\Phi_I)^2\sim (\pounds_V \Phi_I)^2$, which vanishes in equilibrium). In contrast to the main text, there is no explicit dependence on $u^\mu$ in either $\bar p$ or $\Theta$ thanks to the invariance under Lorentz boosts.

Variation of the equilibrium generating functional with respect to the sources is given by
\begin{equation}
\delta\mathbb{W}[g_{\mu\n}, A_\mu, \P^I] = \int d^{d+1}x \sqrt{-g} \left[\frac12 T^{\mu\nu}\delta g_{\mu\nu}+J^\mu\delta A_\mu+O_I\delta\Phi^I \right],
\label{eq:gen_func_va}
\end{equation}
from which we can compute the energy-momentum tensor, the $U(1)$ current and the scalar expectation value, via
\begin{equation}
T^{\mu\nu}_{\rm eq.} = \frac{2}{\sqrt{-g}} \frac{\delta \mathbb{W}}{\delta g_{\mu\n}}\, , \quad J^\mu_{\rm eq.} = \frac{1}{\sqrt{-g}} \frac{\delta \mathbb{W}}{\delta A_\mu},\quad O_I^{\rm eq.} = \frac{1}{\sqrt{-g}} \frac{\delta \mathbb{W}}{\delta \Phi^I}\,.
\label{eq:defsTJ}
\end{equation}
The generating functional is invariant under diffeomorphisms as well as gauge transformations. Vanishing of \eqref{eq:gen_func_va} under the gauge transformation $A_\mu \rightarrow A_\mu + \del_\mu \l(x)$ leads to the Ward identity
\be
\nabla_\mu J^\mu_{\rm eq.} = 0.
\label{eq:Ward_1}
\ee
Similarly, vanishing of \eqref{eq:gen_func_va} under the diffeomorphism $x^\m \rightarrow x^\m + \xi^\m(x)$, combined with \eqref{eq:Ward_1}, yields the Ward identity
\be
\nabla_\m T^{\m\n}_{\rm eq.} = F^{\n}_{\;\;\l} J^\l_{\rm eq.} + O_I^{\rm eq.} \del^\n \Phi^I\, ,
\label{eq:Ward_2}
\ee
where $F_{\m\n} \equiv \del_\m A_\n - \del_\n A_\m$ is the usual anti-symmetric field strength tensor.

The definitions \eqref{eq:thermo_frame} imply that under the variations $g_{\mu\n} \rightarrow g_{\mu\n} + \d g_{\mu\n}$ and $A_\mu \rightarrow A_\mu + \d A_\mu$, we have 
\begin{equation}
\d T = \frac{T}{2} u^\a u^\b \d g_{\a\b}\, , \quad \d\mu = \frac{\mu}{2} u^\a u^\b \d g_{\a\b} + u^\a \d A_\a\, ,
\end{equation}
which together with \eqref{eq:gen_func} and \eqref{eq:defsTJ} lead to
\begin{subequations}\label{eq:eq_currs}
\begin{align}
T^{\mu\n}_{\rm eq.} = &\left[\bar{p}+\Theta \, (\partial\Phi)^2 \right]\! g^{\mu\nu}\! +\! \left[\left(T \f{\del \bar{p}}{\del T} + \mu \frac{\del \bar{p}}{\del \mu} \right) \!+\! \left(T \frac{\del \Theta}{\del T} + \mu \frac{\del \Theta}{\del \mu} \right) (\partial\Phi)^2 \right]\! u^\mu u^\n \!-\!2 \Theta \del^\mu \P^I \del^\n \P_I\, ,\label{eq:eq_stress}\\
J^\mu_{\rm eq.} = & \, \frac{\del \bar{p}}{\del \mu}\, u^\mu + \frac{\del \Theta}{\del \mu}\, (\partial\Phi)^2\, u^\mu\, , \label{eq:eq_current}\\
O^I_{\rm eq.} = &-2\partial_\m\Theta\partial^\m\Phi^I-2\Theta\Box\Phi^I\,.\label{eq:eq_scalar}
\end{align}
\end{subequations}

Note that the thermodynamic relations we have are given by
\begin{subequations}\label{eq:thermodynamics_2}
\begin{align}
T d{\bar s} &= d{\bar \e} - \mu \, d{\bar \rho}, \label{eq:thermod1}\\
T {\bar s} &= {\bar \e} + \bar{p} -\mu {\bar \rho}, \label{eq:thermod2}\\
d{\bar p} &= {\bar s}\, dT + {\bar \rho} \, d\mu , \label{eq:thermod3}
\end{align}
\end{subequations}
where $\bar{\e}, \bar{\rho}$ and $\bar{s}$ are respectively the equilibrium energy density, charge density and entropy density for the fluid, without any correction terms from the scalar fields $\P^I$. In the presence of hydrostatic corrections, we now find the following relations
\begin{subequations}\label{eq:thermodynamics_3}
\begin{align}
dp &= s\, dT + \rho \, d\mu-\Theta \partial^\mu\Phi_I\partial^\nu\Phi^I d g_{\mu\nu} +2\Theta\partial^\mu \Phi^I\, d \partial_\mu \Phi_I, \label{eq:thermod3hs}\\
\e &= -p+T s +\mu \rho\, , \label{eq:thermod2hs}\\
T ds &= d{ \e} - \mu \, d\rho \, - \Theta \partial^\mu\Phi_I\partial^\nu\Phi^I d g_{\mu\nu} +2\Theta\partial^\mu \Phi^I\, d \partial_\mu \Phi_I, \label{eq:thermod1hs}
\end{align}
\end{subequations}
where we have defined
\begin{equation}
\label{thermo_hs}
p=\bar p + \Theta(T,\mu)\, (\partial\Phi)^2\,,\quad s=\bar s +\frac{\del \Theta}{\del T}\, (\partial\Phi)^2,\quad \rho=\bar\rho + \frac{\del \Theta}{\del \mu}\,(\partial\Phi)^2\,.
\end{equation}
Note that when writing variations of $s$, $\rho$ or $\e$, variations of background sources are also generated besides variations of $T$ and $\mu$.
When we go out of equilibrium, we continue to work in the thermodynamic frame. 
This is the underlying frame choice we have worked with in the main text.

\subsection{Entropy current analysis \label{app:entropy}}

Having worked out the hydrostatic corrections, we now turn to the non-hydrostatic sector, which is easiest to determine by computing the divergence of the entropy current and demanding it to be positive semidefinite. We define
\begin{equation}
\Delta\equiv\nabla_\mu s^\mu+\beta_\mu\left(\nabla_\nu T^{\nu\mu} - F^{\mu\nu}J_\nu -O_I\nabla^\mu\Phi^I \right)+\frac{\mu}{T}\,\nabla_\mu J^\mu\, .
\end{equation}
On-shell, $\D$ becomes the divergence of the entropy current and therefore must satisfy $\D \ge 0$. Also, $\b^\m \equiv u^\m/T$.
Using the first law \eqref{eq:thermodynamics_3} and defining $s^\m=su^\m+\tilde s^\m$, we can rewrite $\D$ as
\begin{align}
\Delta=\,\,&\nabla_\m\left[\tilde s^\m+\beta_\n (T^{\mu\nu}+r^{\mu\nu}-\epsilon u^\mu u^\nu)+\frac{\mu}{T}(J^\m-\rho u^\m)\right]-\beta^\mu\nabla_\mu\Phi_I\left[O^I+\nabla_\nu\left(2\Theta\nabla^\nu\Phi^I\right)\right]\nonumber\\
&-\left[T^{\mu\nu}-(\epsilon+p)u^\mu u^\nu-p g^{\mu\nu}+r^{\mu\nu}\right]\nabla_{\mu}\beta_\nu-(J^\mu-\rho u^\mu)\left[\nabla_\mu\left(\frac{\mu}T\right)-F_{\mu\nu}\b^\n\right],
\end{align}
where we have used the notation $r^{\mu\nu}\equiv 2\Theta\nabla^\mu\Phi^I\nabla^\nu\Phi_I$, and the thermodynamic quantities $s,\rho,p$ have been defined in \eqref{thermo_hs} (including $\Theta$-dependent corrections), while $\e=Ts+\mu\rho-p$. The vanishing of entropy production in thermal equilibrium then leads to constitutive relations that include the hydrostatic corrections, \eqref{eq:eq_currs}, as expected. Together with non-hydrostatic corrections that we denote with a tilde, one has
\begin{subequations}\label{eq:eq_currs2}
\begin{align}
T^{\mu\n} &=(\e+p)u^\mu u^\n+p g^{\mu\n} - 2 \Theta \del^\mu \P^I \del^\n \P_I +\tilde T^{\mu\nu}\, ,\label{eq:eq_stress2}\\
J^\mu &=\rho u^\mu +\tilde J^\mu\, , \label{eq:eq_current2}\\
O^I&=-2\partial_\m\Theta\partial^\m\Phi^I-2\Theta\Box\Phi^I+\tilde O^I\,.\label{eq:eq_scalar2}
\end{align}
\end{subequations}
The expression for $\D$ then reads:
\begin{equation}
\label{Delta1}
\Delta=\nabla_\m\left[\tilde s^\m+\tilde T^{\mu\nu} \b_\n+\frac{\mu}{T}\tilde J^\m\right]-\tilde T^{\mu\nu}\nabla_{\mu}\beta_\nu-\tilde J^\mu\left[\nabla_\mu\left(\frac{\mu}T\right)- F_{\mu\nu}\b^\n\right]-\tilde O^I\beta^\mu\nabla_\mu\Phi_I\,,
\end{equation}
from which one can infer the non-hydrostatic contributions to the canonical entropy current:
\begin{equation}
\tilde s^\mu=-\tilde T^{\mu\nu}\b_\n-\frac{\mu}{T}\tilde J^\mu\,.
\end{equation}
Observe that the hydrostatic derivative corrections to $O^I$ only contribute at $\mc{O}(\ell^{3})$, while in principle $\mathcal O(\ell)$ corrections are also allowed. Thus one can anticipate a non-hydrostatic $\mathcal O(\ell)$ contribution to $O^I$, which we indeed find below.

Next, our goal is to write down the non-hydrostatic derivative corrections, which will be of two types: non-dissipative (giving a vanishing contribution to $\Delta$) and dissipative (giving a positive contribution to $\Delta$), ensuring that at the end of the day $\Delta\geq0$. To do so, we need to pick a frame (this does not concern the hydrostatic derivative corrections, which by definition are part of thermodynamic equilibrium), as well as a set of independent derivative corrections. For simplicity, we choose to work in the Landau frame, defined via
\begin{equation}
\label{LFconditions}
u_\mu \tilde T^{\mu\nu}=0\,,\quad u_\mu \tilde J^\mu=0\,,
\end{equation}
which leads to $\tilde s^\mu=-\frac{\mu}{T}\tilde J^\mu$. This simplifies \eqref{Delta1} to
\begin{equation}
\label{Delta2}
T\Delta=-\tilde T^{\mu\nu}\nabla_{\mu}u_\nu-\tilde J^\mu\left[T\nabla_\mu\left(\frac{\mu}T\right)-F_{\mu\nu} u^\nu\right]-\tilde O^I \hat{D}\Phi_I\, ,
\end{equation}
where $\hat{D} \equiv u^\m \nabla_\m$ is the material derivative along the fluid velocity.

We now consider the independent, non-hydrostatic derivative corrections that can contribute to $\tilde T^{\mu\nu}$, $\tilde J^\mu$ and $\tilde O^I$. Recall that the equations of motion are organized in an expansion in powers of $\ell^{2a}$, $a>0$. Terms at ideal order in $T^{\mu\nu}$ and $J^\mu$ contribute to order $a=1$ in the equations of motion, together with terms of order $\ell$ in $O^I$.

The non-hydrostatic terms should by definition vanish on restoring thermal equilibrium. In other words, scalars, vectors and tensors that vanish in thermal equilibrium are the only possible objects that are allowed to appear in the out of equilibrium constitutive relations. Consider first the case of $\mc{O}(\ell^2)$ scalars. Using the formalism laid out in the previous section, it is straightforward to verify that in equilibrium one has
\begin{subequations}
\label{EqbConstraints}
\begin{align}
\N_\l T + T a_\l &= 0\, ,\\
\N_\l \m +\m a_\l - E_\l&= 0\, ,\\
\N_\m u_\n + u_\m a_\n + \f{1}{2} \e_{\m\n\rho\s} u^\rho \O^\s &= 0\, ,
\end{align}
\end{subequations}
where $a^\m \equiv u^\n \N_\n u^\m$ is the acceleration, $E^\m \equiv F^{\m\n} u_\n$ is the electric field, and $\O^\m \equiv \e^{\m\n\rho\s} u_\n \N_\rho u_\s$ is the vorticity vector, all of which are transverse to the fluid velocity. We thus have four $\mc{O}(\ell^2)$ scalars that vanish in thermal equilibrium: $\hat{D}T, \hat{D}\m, \N\cdot u$, and $(\hat{D}\P_I)^2$, where the last one vanishes in equilibrium because of the condition $\pounds_V \Phi^I = 0$. Two of these scalars, $\hat{D}T$ and $\hat{D}\m$, can be eliminated in favour of $\N\cdot u$ using the background equations of motion, leaving behind only two independent scalars at $\mc{O}(\ell^2)$:
\be
\label{l2scalars}
\N\cdot u\, , \quad(\hat{D}\Phi_I)^2\, .
\ee

Next, consider $\mathcal{O}(\ell^2)$ transverse vectors, which satisfy the Landau frame condition \eqref{LFconditions} and can therefore appear in the constitutive relation for $\tilde{J}^\m$. Using the equilibrium conditions \eqref{EqbConstraints}, one finds three transverse vectors at $\mc{O}(\ell^2)$ that vanish in equilibrium: $\mathds{P}^{\m\n}(\N_\n T+ Ta_\n), \mathds{P}^{\m\n} \left(T\N_\n\left(\f{\m}{T}\right) -E_\n\right)$ and $\mathds{P}^{\m\n} \N_\n \P_I \hat{D}\P^I$, where $\mathds{P}_{\m\n} \equiv g_{\m\n} + u_\m u_\n$ is the transverse projector. Of these, we can eliminate the first in favour of the rest using the background equations of motion, leading to two independent transverse vectors at $\mc{O}(\ell^2)$:
\be
\label{l2vectors}
\mathds{P}^{\m\n} \left(T\N_\n\left(\f{\m}{T}\right) -E_\n\right) , \quad \mathds{P}^{\m\n} \N_\n \P_I \hat{D}\P^I.
\ee

Let us now count the possibilities for symmetric tensors at $\mc{O}(\ell^2)$ which can appear in the constitutive relation for $\tilde{T}^{\m\n}$. These have to be transverse to satisfy the Landau frame condition \eqref{LFconditions}. Further, one can decompose a symmetric tensor into a traceless part and a trace. We have already enumerated independent $\mc{O}(\ell^2)$ scalars in \eqref{l2scalars}, which can appear in the trace. For the symmetric transverse traceless part, there is only one possibility at $\mc{O}(\ell^2)$, given by the shear tensor,
\be
\label{l2tensor}
\s^{\m\n} \equiv \mathds{P}^{\m\a} \mathds{P}^{\n\b} \left(\N_\a u_\b + \N_\b u_\a - \f{2}{d} \, g_{\a\b}  \N\cdot u \right).
\ee

Finally, let us consider the possibilities for $\mc{O}(\ell)$ as well as $\mc{O}(\ell^3)$ scalars, which can appear in the constitutive relation for $\tilde{O}_I$. At $\mc{O}(\ell)$, there is a single possibility: 
\be
\label{lscalar}
\hat{D}\P_I\,.
\ee

At $\mc{O}(\ell^3)$, one has six possible independent contributions: 
\be
\label{l3scalars}
\begin{split}
&\N\cdot u \, \hat{D}\P_I\, , \quad (\hat{D}\P_J)^2 \hat{D}\P_I\, ,\quad (\N_\m\P_J)^2 \hat{D}\P_I\, , \quad \hat{D}^2 \P_I\, ,\\
&\N_\m \P_I \N^\m \P^J \hat{D}\P_J\, ,\quad  \mathds{P}^{\m\n}\left(T\N_\n\left(\f{\m}{T}\right) - E_\n\right) \N_\m \P_I\,. 
\end{split}
\ee

Using the above results, we find it convenient to parametrize the non-hydrostatic corrections in the constitutive relations \eqref{eq:eq_currs2} as\footnote{Note that $\nabla_{(\alpha} u_{\beta)} \equiv \frac{1}{2}\left(\nabla_\alpha u_\beta + \nabla_\beta u_\alpha\right)$.}
\begin{subequations}\label{eq:nh_corrs_1}
\begin{align}
\tilde T^{\mu\nu} &=-\eta^{\mu\nu\alpha\beta}\nabla_{(\alpha} u_{\beta)}-\gamma^{\mu\nu I}\hat{D}\Phi_I-\lambda^{\mu\nu\alpha}X_\alpha\,,\quad X_\alpha\equiv T\nabla_\alpha\left(\frac{\mu}T\right)- E_\alpha\\
\tilde J^{\mu} &=-\Sigma^{\mu\nu} X_\nu- \lambda^{\alpha\beta\mu}\nabla_{(\alpha}u_{\beta)}-\Xi^{\mu I}\hat{D}\Phi_I,\\
\tilde O^{I} &=-h^{IJ}\hat{D}\Phi_J+ \Xi^{I\mu}X_\mu+ \gamma^{\alpha\beta I}\nabla_{(\alpha}u_{\beta)}\, ,
\end{align}
\end{subequations} 
where we have already imposed Onsager relations to fix the relative signs of cross-terms. From this, we immediately note that $\gamma$ and $\Xi$ will only include non-dissipative cross-terms as their contributions will vanish from $\Delta$. We should now expand all the tensors in \eqref{eq:nh_corrs_1} order by order in $\ell$, allowing only terms compatible with the Landau frame conditions \eqref{LFconditions} and included in the set of independent terms we just enumerated in \eqref{l2scalars}, \eqref{l2vectors}, \eqref{l2tensor}, \eqref{lscalar} and \eqref{l3scalars}.

The independent derivative corrections are
\begin{subequations}\label{eq:nh_corrs_2}
\begin{align}
\eta^{\mu\nu\alpha\beta} &=\eta \left(\mathds{P}^{\mu\alpha}\mathds{P}^{\nu\beta}+\mathds{P}^{\mu\beta}\mathds{P}^{\nu\alpha}-\frac{2}{d}\,\mathds{P}^{\mu\nu}\mathds{P}^{\alpha\beta}\right)+\zeta \, \mathds{P}^{\mu\nu} \mathds{P}^{\alpha\beta}+\mc{O}(\ell^2),\\
\lambda^{\mu\nu\alpha} &=\mc{O}(\ell^2),\\
\gamma^{\mu\nu I}&=\tilde\zeta\, \mathds{P}^{\mu\nu} \hat{D}\Phi^I+\mc{O}(\ell^3),\\
\Sigma^{\mu\nu}&=\sigma_o \, \mathds{P}^{\mu\nu}+\mc{O}(\ell^2),\\
\Xi^{\mu I}&=\xi \, \mathds{P}^{\mu\nu}\nabla_\nu\Phi^I+\mc{O}(\ell^3),\\
h^{IJ}&=\delta^{IJ}\left(\gamma+\gamma_1 (\nabla\Phi)^2+\gamma_2 (\hat{D}\Phi)^2+\gamma_4 \hat{D}\right)+\gamma_3 \nabla^\mu\Phi^I\nabla_\mu\Phi^J+\mc{O}(\ell^5).
\end{align}
\end{subequations} 
We did not include a $\delta^{IJ}\nabla\cdot u$ term in $h^{IJ}$ as this term is already included in $\gamma^{\mu\nu I}$, or a $\hat{D}\Phi^I \hat{D}\Phi^J$ term as it would give the same contribution in $\tilde O^I$ as the $\gamma_2$ term. If we chose a different frame, then we would need to write the derivative corrections in terms of $\nabla_{(\mu} \beta_{\nu)}$ instead of $\nabla_{(\mu} u_{\nu)}$. The $\gamma_4$ term has the differential operator $\hat{D}\equiv u^\mu \nabla_\mu$ acting on $\hat{D}\Phi_I$. Such differential operators will also appear at higher orders in $\ell$ in other tensors, or in other frames to the order we are working. 

If one wants to impose conformal symmetry, then the trace of the stress-energy tensor must vanish,\footnote{Since the $O_I$'s are marginal operators.} $T^\mu_\mu=0$, which in turn imposes $\zeta=\tilde\zeta=0$, as well as, from \eqref{eq:eq_currs},
\begin{equation}
\label{conformalsym}
\bar p = T^{d+1}f\left(\frac{T}\mu\right),\quad \Theta=T^{d-1}g\left(\frac{T}\mu\right),
\end{equation}
for some undetermined functions $f$ and $g$ of $T/\mu$.

Using \eqref{eq:nh_corrs_1}, the divergence of the entropy current reads:
\begin{equation}
T \Delta=\eta^{\mu\nu\alpha\beta}\nabla_{\mu}u_\nu\nabla_{\alpha}u_\beta+\Sigma^{\mu\nu}X_{\mu}X_\nu+h^{IJ}\hat{D}\Phi_I \hat{D}\Phi_J\, ,
\end{equation}
which upon inserting \eqref{eq:nh_corrs_2} becomes
\begin{equation}
\begin{split}
T\Delta&=\gamma\left[\hat{D}\Phi_I+\frac{\hat{D}\Phi_I}{2\gamma}\left(\gamma_1 (\nabla\Phi)^2 +\gamma_2 (\hat{D}\Phi)^2\right)+\frac{\gamma_4}{2\gamma} \hat{D}^2\Phi_I+\frac{\gamma_3}{2\gamma} \nabla_\mu\Phi_I\nabla^\mu\Phi^J \hat{D}\Phi_J\right]^2\\
&\quad+\f{\eta}{2} \, \sigma_{\mu\nu}\sigma^{\mu\nu}+\zeta(\nabla\cdot u)^2+\sigma_o \,(\mathds{P}^{\mu\nu}X_\nu)^2 +\mc{O}(\ell^6)\, .
\end{split}
\end{equation}
In the expression above we are actually neglecting the $\mc{O}(\ell^6)$ terms produced by expanding the square in the first line, which lie beyond the derivative order to which we are working. 
The requirement $\Delta\geq0$ therefore results into the following constraints:
\begin{equation}
\eta,\,\zeta,\,\sigma_o,\,\gamma\geq0\,.
\end{equation}

We may now compute the thermoelectric conductivities, which read
\begin{subequations}\label{eq:conduc_rel}
\begin{align}
\sigma(\omega)&=\sigma_o+\frac{(\rho-\ell^2\xi)^2}{(Ts+\mu \rho+\ell^2\gamma_4)}\frac1{\Gamma-i\omega}\, ,\\
\alpha(\omega)&=-\frac{\mu}{T}\sigma_o+\frac{(\rho-\ell^2\xi)(Ts+\ell^2\mu\xi)}{T(Ts+\mu \rho+\ell^2\gamma_4)}\frac1{\Gamma-i\omega}\, ,\\
\bar\kappa(\omega)&=\frac{\mu^2}{T}\sigma_o+\frac{(Ts+\ell^2\mu\xi)^2}{T(Ts+\mu \rho+\ell^2\gamma_4)}\frac1{\Gamma-i\omega}\, ,
\end{align}
\end{subequations} 
where 
\begin{equation}
\Gamma\equiv\ell^2\frac{\gamma+\ell^2(d\gamma_1+\gamma_3)}{Ts+\mu \rho+\ell^2\g_4}+\mc{O}(\ell^6)\,.
\end{equation}
Comparing these expressions to \eqref{acsigma}-\eqref{ackappa} and using \eqref{Lorentz}, we identify
\begin{equation}
\l_v = \g_4 \, , \quad \l_\rho = - \xi\, , \quad \l_s = \f{\m}{T}\xi\, .
\end{equation}
Thus the non-dissipative, non-hydrostatic coefficients $\l_{\rho,s}$ map to $\xi$, which is of the same nature, while the dissipative coefficients $\l_v$ and $\g_4$ map on to one-another.

\section{Comparison with holographic models}
\label{app:comp_holo}
The effective theory we have developed should match on to holographic models of metallic phases in the regime of weak breaking of translation invariance. More specifically, thanks to our assumption that the scalars enjoy a shift symmetry, the holographic model we wish to compare to is given by \cite{Andrade:2013gsa} (setting $16\pi G_N=1$)
\begin{equation}
\label{holoaction}
S=\int d^{d+2}x\sqrt{-g}\left[R-2\Lambda-\frac14 F^2-\frac12 \nabla_A\Phi_I\nabla^A\Phi^I\right].
\end{equation}
Here capital Latin indices $A,B=0,1,\ldots d+1$ run over the bulk spacetime coordinates, while Greek lowercase indices run over the boundary spacetime coordinates, $\mu,\nu=0,1,\ldots d$, lowercase Latin indices $i,j=1,\ldots d$ run over the boundary spatial coordinates, and capital Latin internal indices $I,J=1,\ldots d$ run over the number of scalars, which for simplicity we take to be the same as the number of boundary spatial dimensions. We also raise and lower capital Latin internal indices with $\delta^{IJ}$, although one could choose a non-trivial target space metric if desired. 
Assuming that the background value for the scalars is
\begin{equation}
\Phi^I=\ell \delta^{I}_i x^i\,,
\end{equation}
a planar, isotropic black hole solution can be found \cite{Bardoux:2012aw,Andrade:2013gsa}, with
\begin{equation}
ds^2=-r^2f(r)dt^2+\frac{dr^2}{r^2f(r)}+r^2d\vec x^2, 
\end{equation}
together with
\begin{equation*}
f(r)=\left(1-\frac{\ell^2r^2}{2(d-1)}\right)\left(1-\frac{r_0^{d+1}}{r^{d+1}}\right)-\frac{(d-1)\mu^2r_0^{d-1} }{2d r^{d+1}}\left(1-\frac{r_0^{d-1}}{r^{d-1}}\right),\quad A_t=\mu\left(1-\frac{r_0^{d-1}}{r^{d-1}}\right).
\end{equation*}
In these coordinates, the AdS$_{d+2}$ boundary is located at $r=\infty$ and the horizon at $r=r_0$. $\mu$ is the chemical potential, and the parameter $\ell$ controls the strength of momentum relaxation. In the regime where $\ell$ is small compared to $T,\mu$, a Drude-like long-lived excitation is found in the linear, thermoelectric response, \cite{Andrade:2013gsa,Davison:2014lua,Davison:2015bea,Blake:2015epa,Blake:2015hxa}.

Before comparing the hydrodynamic regime with our effective theory, it is instructive to check the thermodynamic properties. In \cite{Andrade:2013gsa}, it was found for $d=2$ that
\begin{equation}
\rho=\mu r_0\,,\quad s=4\pi r_0^2\,,\quad 4\pi T=3r_0-\frac{\ell^2}{2r_0}-\frac{\mu^2}{4 r_0}\,,
\end{equation} 
\begin{equation}
 \epsilon=2r_0^3\left(1-\frac{\ell^2}{2r_0^2}+\frac{\mu^2}{4r_0^2}\right),\quad p=r_0^3+\frac12\ell^2 r_0+\frac{1}{4}\mu^2r_0\,.
\end{equation} 
By substituting the expression for $r_0(T,\mu,\ell)$ into the various quantities and expanding to $\mathcal O(\ell^2)$, it is straightforward to identify the static susceptibility $\Theta$ introduced in \eqref{eq:gen_func} as $\Theta=r_0/2$. More generally, by using the generic holographic prescription for computing the free energy, and the fact that $s=4\pi r_0^{d-1}$ and $\rho=(d-1)\mu r_0^{d-1}$, we find that
\begin{equation}
\label{Thetaholstate}
\Theta=\frac{r_0^{d-1}}{2(d-1)}\,.
\end{equation}
This continues to hold for the case $d=3$ investigated in \cite{Blake:2015epa}, with care being taken about the difference in normalization of the $F^2$ term in their action compared to \eqref{holoaction}.

We can do even better and verify that the expressions derived for the stress-tensor in \cite{Blake:2015epa,Blake:2015hxa} match in the hydrostatic sector to \eqref{eq:eq_currs}. There, it was found that
\begin{equation}
T^{\mu\nu}=(d+1)b u^\mu u^\nu+b \mathds{P}^{\mu\nu}-\eta \sigma^{\mu\nu}-\eta_\Phi \Phi^{\mu\nu},
\end{equation}
where $b$ is related to the horizon radius and the charge of the black hole, and the specific expressions for $\eta_\Phi$ in $d=2,3$ are compatible with $\eta_\Phi=r_0^{d-1}/(d-1)=2\Theta$. Moreover,
\begin{equation}
\Phi^{\mu\nu}=\mathds{P}^{\mu\alpha} \mathds{P}^{\nu\beta}\nabla_\alpha\Phi_I\nabla_\beta\Phi^I-\frac{1}{d}\,\mathds{P}^{\mu\nu}\mathds{P}^{\alpha\beta}\nabla_\alpha\Phi_I\nabla_\beta\Phi^I,
\end{equation}
which can be expanded as
\begin{equation}
\Phi^{\mu\nu}=\nabla^\mu\Phi_I\nabla^\nu\Phi^I+2\hat{D}\Phi_I u^{(\mu}\nabla^{\nu)}\Phi^I+(\hat{D}\Phi_I)^2u^\mu u^\nu-\frac{1}{d} \, \mathds{P}^{\mu\nu}(\hat{D}\Phi_I)^2-\frac{1}{d}\,\mathds{P}^{\mu\nu}(\nabla\Phi_I)^2\,.
\end{equation}
Of these terms, the first and the last do not vanish over equilibrium configurations, and can be mapped to the expression for the stress-energy tensor with hydrostatic corrections in \eqref{eq:eq_currs}. This requires one to be careful as the parameter $r_0$ used in \cite{Blake:2015epa,Blake:2015hxa} picks up $\mathcal O(\ell^2)$ corrections when re-expressed in terms of $T$ and $\mu$. The hydrostatic part of the current can be similarly matched to \eqref{eq:eq_currs}, being equally careful with the parameter $q$ used in \cite{Blake:2015epa,Blake:2015hxa}. On  the other hand, there does not seem to be a way to obtain the hydrostatic corrections to $O_I$ from the expressions derived in \cite{Blake:2015epa,Blake:2015hxa}, where the scalar vev only has non-hydrostatic corrections that vanish in equilibrium. Refs.~\cite{Blake:2015epa,Blake:2015hxa} explicitly mention neglecting a number of terms which do not contribute to linear response, such as $u^\mu u^\nu\nabla_\alpha\Phi_I\nabla_\beta\Phi_J$, so it is plausible that the $\nabla(\Theta\nabla\Phi)$ term we find was also neglected. 

For similar reasons, it is difficult to make a precise comparison with \cite{Blake:2015epa,Blake:2015hxa} in the non-hydrostatic sector. However, as we now discuss, our results do match for the thermoelectric conductivities. Refs.~\cite{Blake:2015epa,Blake:2015hxa} found equivalent results to \cite{Davison:2015bea}, where in $d=2$ one has
\begin{subequations}
\begin{align}
\sigma\left(\omega\right)&=\frac{\frac{\rho^2}{\epsilon+p}+\Gamma\left(1-\sigma_o+\lambda\mu^2\right)+O(\Gamma^2,
\omega\Gamma,\omega^2)}{\Gamma-i\omega}+\sigma_o+O(\omega,\Gamma)\, ,\label{acsigmaholo}\\
\alpha\left(\omega\right)&=\frac{\frac{\rho s}{\epsilon+p}+\Gamma\left(\frac{\mu}{T}\sigma_o+4\pi \rho\lambda\right)+O(\Gamma^2,
\omega\Gamma,\omega^2)}{\Gamma-i\omega}-\frac{\mu}{T}\sigma_o+O(\omega,\Gamma)\, ,\label{acalphaholo}\\
\bar{\kappa}\left(\omega\right)&=\frac{\frac{s^2T}{\epsilon+p}+\Gamma\left(-\frac{\mu^2}{T}\sigma_o+4\pi s T\lambda\right)+O(\Gamma^2,
\omega\Gamma,\omega^2)}{\Gamma-i\omega}+\frac{\mu^2}{T}\sigma_o+O(\omega,\Gamma)\, ,\label{ackappaholo}
\end{align}
\end{subequations}
together with 
\begin{equation}
\label{eq:GammaExp2}
\begin{aligned}
\Gamma&=\frac{s \ell^2}{4\pi\left(\epsilon+p\right)}\left[1+\lambda \ell^2+O(\ell^4)\right],\\
\lambda&=\frac{\sqrt{3}\pi-9\log3}{96\pi^2T^2}+\frac{9\mu^2\left(\log3-2\right)}{256\pi^4 T^4}-\frac{9\mu^4\left(42\log3+5\sqrt{3}\pi-132\right)}{32768\pi^6 T^6}+O\left(\frac{\mu^6}{T^8}\right),
\end{aligned}
\end{equation}
and
\begin{equation}
\label{eq:rnads4sigmaq}
\sigma_o=\left(\frac{Ts}{3\epsilon/2}\right)^2\Biggr|_{\ell=0}=\left.\left(\frac{3-\frac{\mu^2}{4r_0^2}}{3\left(1+\frac{\mu^2}{4r_0^2}\right)}\right)^2\right|_{\ell=0}.
\end{equation}
Importantly, in \eqref{acsigmaholo}-\eqref{ackappaholo}, the thermodynamic quantities appearing include the hydrostatic $\mathcal O(\ell^2)$ corrections.
Working to $\mathcal O(\ell^4)$, these expressions can be matched to \eqref{acsigma}-\eqref{ackappa} as described in the main text around \eqref{matchholo}.

\bibliography{biblio}

\end{document}